# Quantum confinement effect in Sb thin films

Anuradha Wijesinghe[1,*], Yongxi Ou[2,3,*], Anjali Rathore[1], Chandima Edirisinghe[1], Pradip Adhikari[1], An-Hsi Chen[4], Dustin Gilbert[5], Anthony Richardella[2], Nitin Samarth[2], Joon Sue Lee[1,†]

[1]Department of Physics and Astronomy, University of Tennessee, Knoxville, TN
[2]Department of Physics and Astronomy, Pennsylvania State University, State College, PA
[3]School of Integrated Circuits, Nanjing University, Suzhou, Jiangsu, 215163, China
[4]Oak Ridge National Laboratory, Oak Ridge, TN
[5]Material Science and Engineering Department, University of Tennessee, Knoxville, TN
[*]These authors contributed equally to this work.
[†]Corresponding author e-mail: jslee@utk.edu

**ABSTRACT**

Antimony (Sb), an element with strong spin-orbit coupling, is predicted to undergo a topological phase transition from a topological semimetal to a topological insulator as its dimensionality approaches the two-dimensional limit, driven by the quantum confinement effect. In this study, we investigate this transition in Sb thin films grown by molecular beam epitaxy, employing electrical transport measurements and angle-resolved photoemission spectroscopy (ARPES). Electrical transport measurements revealed signatures of a modified electronic band structure, including a Hall response with multiple carrier types, a decreasing carrier concentration, and a transition in the curvature of the longitudinal resistance from quadratic to linear with decreasing film thickness. Temperature-dependent magnetoresistance further showed weak antilocalization below 16 K, indicating strong spin-orbit coupling and suggesting the presence of non-trivial topological states. Analysis of the WAL characteristics revealed a single coherent conducting channel and a thickness-dependent change in the phase decoherence mechanism. Complementary ARPES measurements confirmed that reducing the film thickness lifts the conduction band at the M-point, consistent with the emergence of a band gap. These findings support theoretical predictions of a thickness-dependent band structure evolution driven by the quantum confinement effect, providing a foundation for further exploration of topological phase transitions in Sb as well as $Bi_{1-x}Sb_x$. The realization of an elemental topological material with simplified stoichiometry and semiconductor compatibility presents a promising avenue for next-generation hybrid systems and applications in spintronics and quantum technologies.

# I. INTRODUCTION

Antimony (Sb), a group VA (group 15) element, has long been recognized as a semimetal with remarkable properties such as high carrier mobility, tuneable band gap, and fast optical and thermoelectric responses, making it a subject of significant research interest over the years [1–4]. In addition, Sb exhibits strong spin-orbit coupling and has served as a key component in various topological materials, including $Bi_{1-x}Sb_x$ and $Sb_2Te_3$, which have played pivotal roles in the early development of the field [5] [6]. More recently, Sb has garnered attention for its intrinsic topological nature, exhibiting phases such as topological semimetal (TSM), three-dimensional (3D) topological insulator (TI), and the quantum spin Hall (QSH) state, all having topologically non-trivial band structures [7–10]. In its monolayer form, antimonene has emerged as a promising two-dimensional material due to its ambient stability, high carrier mobility, and potential for thermoelectric and ferroelectric applications. It has also been predicted to host a robust QSH state, further enhancing interest in Sb-based low-dimensional systems [11–14]. Furthermore, Sb thin films can undergo a topological phase transition from a TSM to a 3D TI when the quantum confinement effect comes into play as the material approaches the two-dimensional (2D) limit [7,15].

In a semimetal, quantum confinement modifies the electronic band structure, leading to an increase in the band gap as the thickness decreases. This arises from the opposite signs of the effective masses of electrons and holes in the conduction and valence bands, respectively. Considering the band dispersion of Sb along the Γ-M direction, there exists a direct band gap at the Γ point and an indirect negative band gap between the lowest point of the conduction band and the highest point of the valence band. Due to the quantum confinement effect, both band gaps tend to increase, and at a certain thickness, the indirect negative band gap vanishes, leaving a continuous band gap throughout the Γ-M direction. During this process, the topology of the bands remains unchanged, and as a result, Sb is predicted to undergo a topological phase transition from a TSM to a 3D TI (TI) when the thickness decreases below 7.8 nm [7].

TIs are unique materials characterized by insulating bulk states and conducting surface states, where the electron spin is locked to its momentum, resulting in a perpendicular spin-momentum relationship. This property protects the surface states from elastic backscattering, enabling efficient spin-charge conversion, making TIs highly promising for spintronics applications [16,17]. TSMs are distinguished by band crossings between the conduction and valence bands near the Fermi level. These band crossing points are topologically protected, persisting despite perturbations [2]. TSMs offer a range of properties such as superconductivity, semiconducting behaviour, and other semimetal characteristics, and this versatility has led to applications in quantum computing, superconducting device fabrication, energy harvesting, and topological quantum devices [18–20]. Sb stands out due to its potential as a TI and a TSM. Moreover, as an elemental material, the simple stoichiometry reduces defect formation and enhances compatibility with other materials, making Sb highly relevant for the development of advanced hybrid systems and the ongoing exploration of topological phases. Experimentally realizing the predicted phase transition in Sb could thus open new pathways for hybrid spintronics and quantum device applications.

Despite theoretical predictions, experimentally verifying the topological phase transition in Sb thin films remains challenging. Transport measurements have yet to provide conclusive evidence, as bulk conduction can obscure the contribution of topological surface states (TSS). Additionally, defects introduced during growth further complicate the interpretation of transport results [21]. This was evident in previous studies where surface transport was demonstrated but confirmation of the topological phase transition was not clearly seen due to the contribution from bulk bands and limited surface coverage at low thicknesses [22]. In this work, we carefully investigated this transition thickness by combining electrical transport measurements and angle-resolved photoemission spectroscopy (ARPES), which enabled to indirectly and directly probe changes in the band structure of Sb as its thickness approached the range predicted for the topological phase transition.

Sb thin films with minimized defects were successfully grown using molecular beam epitaxy (MBE) along the desired (001) hexagonal plane, which was confirmed by structural and morphological characterizations. Hall measurements demonstrated the coexistence of multiple carrier types, possibly originating from the conduction band, valence band, and surface states, with their concentrations varying as a function of thickness. Longitudinal magnetoresistance (MR) in all Sb thin films revealed weak antilocalization, indicative of strong spin-orbit coupling in Sb and the possible presence of TSS. ARPES showed clear changes in band dispersion with thickness, providing direct evidence for a thickness-driven band gap opening. Consistent results were obtained between transport measurements and ARPES as the Sb thickness varied. This represents a promising step towards realizing the theoretically predicted transition from a TSM to a TI.

## II. EXPERIMENTAL

Two batches of Sb films were prepared: one on GaSb(111)B substrates and another on sapphire(0001) substrates, each grown in separate MBE chambers. Sb films grown on GaSb(111)B were used for electrical characterization, while Sb films grown on sapphire(0001) substrates were used for ARPES measurements. During the growth, the substrate temperature was monitored by a pyrometer. In the following growth procedure, the pyrometer temperature $T_p$ is stated when the substrate temperature is above 250°C, where the pyrometer is functional. For instances near or below $T_p$ = 250°C, manipulator temperature $T_m$ is stated. GaSb(111)B substrates were outgassed at $T_m$ = 500°C in a separate outgassing chamber and transferred to III-V MBE chamber, where they were heated to remove the native oxide in the presence of $Sb_2$ flux. The oxide desorption was identified while heating up the substrates by the appearance of streaky (8 x 1) reflection high-energy electron diffraction (RHEED) pattern, at which point the pyrometer was calibrated as $T_p$ = 540°C. Then, the substrates were annealed at $T_p$ = 545 °C for 10 minutes to completely remove the oxide layer, followed by cooling to $T_p$ = 480°C for the growth of a 20-nm-thick GaSb buffer layer, which smooths out any surface roughness resulting from oxide desorption. Next, the samples were cooled down to room temperature ($T_m$ =20°C) and left for 8 hours or more to ensure a uniform temperature distribution. Based on previous growth trials indicating that a seed layer enhances the surface morphology of Sb thin films, a 2-nm-thick seed layer was first deposited at 20°C. This low-temperature growth led to the disappearance of the RHEED pattern from the GaSb layer, as epitaxial growth is not favoured at such temperatures. After the deposition of the seed layer, the substrate temperature was gradually increased to $T_m$ = 300°C at a steady ramp rate of 20°C/min. When the RHEED pattern reappeared at approximately 150°C, the Sb shutter was opened to initiate film growth. The growth continued as $T_m$ increased to 300°C and proceeded at that temperature until the intended film thickness was achieved. Samples of varying thicknesses (5.1 nm to 13.2 nm) were grown, with the thickness determined by X-ray reflectivity (XRR), as summarised in Table 1. The XRR fitting done for the 13.2-nm-thick sample is shown in Fig. S1 in the supplementary material.

Table 1. Thickness of the Sb films grown on GaSb(111)B and sapphire(0001) substrates, determined using XRR.

| Sample | Sb layer | Buffer layer | Substrate |
|---|---|---|---|
| A | 5.1 nm | GaSb, 20 nm | GaSb(111)B |
| B | 6.6 nm | GaSb, 20 nm | GaSb(111)B |
| C | 8.7 nm | GaSb, 20 nm | GaSb(111)B |
| D | 8.9 nm | GaSb, 20 nm | GaSb(111)B |
| E | 9.9 nm | GaSb, 20 nm | GaSb(111)B |
| F | 13.2 nm | GaSb, 20 nm | GaSb(111)B |
| G | 6 nm | $(Bi,Sb)_2Te_3$, 2 nm | Sapphire(0001) |
| H | 15 nm | $(Bi,Sb)_2Te_3$, 2 nm | Sapphire(0001) |

Two samples for ARPES measurements were grown on sapphire substrate with thicknesses of 6 nm and 15 nm. After outgassing the sapphire substrates, a 2 nm $(Bi,Sb)_2Te_3$ layer was grown as a buffer layer. As with the previous batch of samples, the growth of the Sb films began at room temperature. The substrate temperature was then gradually increased to 70°C, and the film growth was continued until the desired thickness was achieved. Finally, the samples were *in-vacuo* transferred to an interconnected ARPES chamber. ARPES measurements were carried out at 77 K using the 21.2 eV spectral line from a helium plasma lamp and detected by a Scienta Omicron DA 30 L analyser with an energy resolution of 6 meV and the band dispersion along two different directions, ($\overline{\Gamma K}$) and ($\overline{\Gamma M}$). A clear shift in the bands was observed in the $\overline{\Gamma K}$ direction with the change in sample thickness.

All the samples were characterized by atomic force microscopy (AFM) using a Cypher S AFM system to examine their morphology, which revealed films with triangular features for the Sb films grown on GaSb(111)B, while the samples grown on sapphire did not display a clear, distinctive triangular morphology. X-ray diffraction (XRD) was performed using a Malvern Panalytical X'pert3 Materials Research Diffractometer to verify the growth direction of the Sb films, confirming that the Sb films grew along the (001) hexagonal direction on both GaSb(111)B and $(Bi,Sb)_2Te_3$/sapphire substrates. The electrical transport properties of the films were measured using a cryostat with a 9 T magnet down to 4 K. The resistance of the samples was assessed using a Hall bar pattern scratched onto the surface, with the longitudinal resistance, $R_{xx}$ along the $[\overline{2}11]$ crystal orientation, was measured as a function of temperature, *T* and the perpendicular magnetic field, *B*. Low-temperature MR measurements revealed weak antilocalisation, while Hall measurements indicated mixed carrier transport. Additionally, these measurements provided insights into the phase decoherence mechanism and the potential presence of non-trivial conduction, which will be discussed further in the Results and Discussion section.

## III. RESULTS AND DISCUSSION

### A. Structural Characterization

XRD performed on the 13.2-nm-thick Sb film grown on GaSb and the 15-nm-thick Sb film on sapphire clearly shows peaks of Sb(003) and Sb(006), which verify that the growth of Sb on both GaSb(111) and sapphire substrates is along the $(001)_H$ direction of hexagonal Sb as shown in Fig. 1(a). The Sb peaks are prominent and hence indicate a decent film growth on the selected substrates. The AFM images relating to the Sb film grown on sapphire showed grain-like features while Sb on GaSb(111) comprised of triangular-shaped features, which resembles the hexagonal lattice structure of Sb in the $(001)_H$ plane (Figs. 1(b) and 1(c) respectively), which is the most common crystal structure followed by group-VA solid elements and comprises bilayers of thickness 3.75 Å stacked along the [111] direction [23]. The height of the triangular features of Sb on GaSb was determined using a line profile drawn on the AFM image in Fig. 1(b). This line profile showed that the average height of a triangular feature is 3.78 Å (Fig. 1(d)), which is about the height of a bilayer of Sb along the (111) plane, as illustrated in Fig. 1(e).

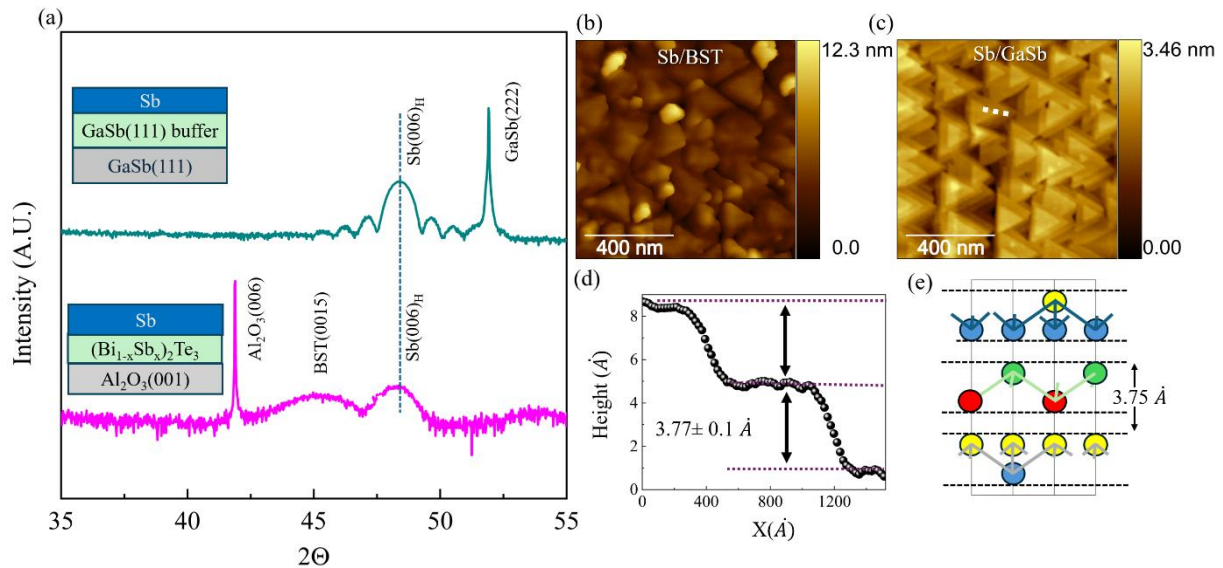

Figure 1. (a) XRD of Sb grown on GaSb and sapphire substrates confirming the growth of Sb along the $(001)_H$ direction. Insets show the cross-sections of the films grown using GaSb(111)B and sapphire substrates respectively. (b) and (c) AFM images of Sb on sapphire and GaSb respectively showing different surface morphologies of Sb. (d) Height profile of a line drawn over a triangular feature in (c) (line is marked as a dotted white line in the AFM image). (e) Cross-section of hexagonal Sb crystal structure comprising of bilayers showing the atomic structure and bilayer thickness.

### B. Electrical Characterization with Thickness Variation

To study the electrical properties of the Sb films with varying thicknesses, a non-lithographic technique was used to fabricate Hall bars by simply scratching the Sb film using a mechanically controlled needle. The electrical transport measurement setup is shown in the inset of Fig. 2. Temperature dependence of longitudinal sheet resistance reveals semimetallic nature of the Sb thin films, grown on GaSb(111)B substrates, except for the thinnest film (5.1 nm), as shown in Fig. 2. As the temperature decreases, carriers in GaSb substrate start to freeze below 150 K, making it highly insulating, and hence the resistance behaviour below this temperature emerges mostly from the Sb layer. Below 80 K, in Sb thin films equal to or greater than 6.6 nm, the resistance shows a metallic behaviour with a positive slope, decreasing linearly down to around 40 K, beyond which it begins to saturate. With the decrease in thickness down to 6.6 nm, the resistance tends to increase, and the 5.1-nm-thick film shows an insulating behaviour with a negative slope. This result of temperature dependence with varying Sb film thickness is consistent with the theoretical proposal of shifting of conduction and valence bands to open a gap with a decrease in thickness.

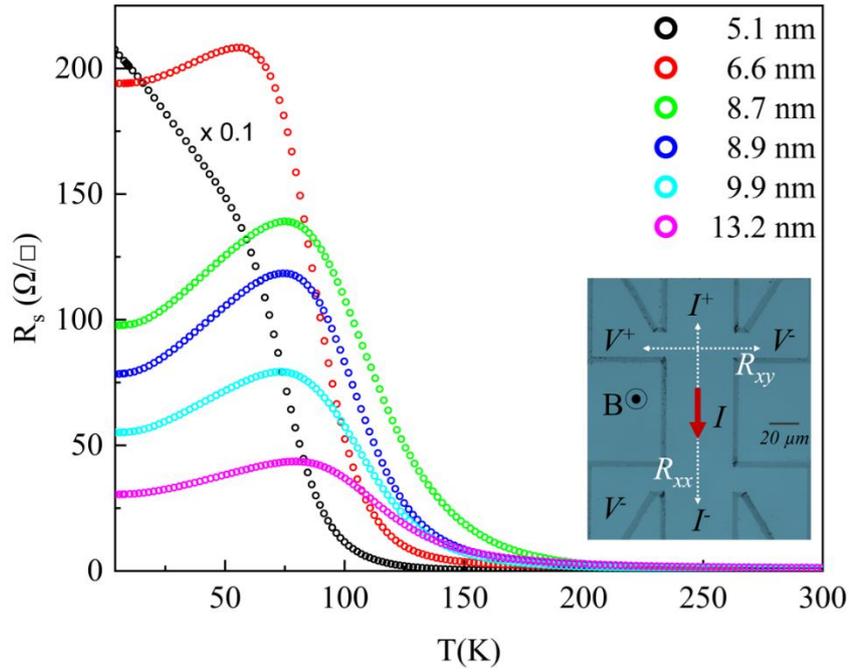

Figure 2. Sheet resistance vs temperature plots for the films showing metallic behaviour of Sb on GaSb films except for the lowest thickness film showing an insulating nature at low temperatures. *Inset* shows an optical microscopic image of a scratched Hall bar showing the configuration setup to determine $R_{xx}$ and $R_{xy}$.

Longitudinal and transverse MRs ($R_{xx}$ and $R_{xy}$) were obtained with perpendicular magnetic field, $B$ at finite temperatures at 4 K (Figs. 3(a) and 3(b)). In Fig. 3(a), the nature of the longitudinal MR was analysed by applying a power-law fitting, where the exponent γ in the $B^\gamma$ relationship provided a quantitative measure of the curvature of the MR plots. The exponent decreased from 1.38 in the 13.2 nm film (close to quadratic behaviour) to approximately 1.0 in the 8.7 nm film (linear), and further to 0.57 in the 5.1 nm film (sub-linear), as shown in the inset of Fig. 3(a), indicating a clear transition to sub-linear MR with decreasing thickness. A similar behaviour in the longitudinal MR curves has been observed in TI thin films such as $Bi_2Se_3$, where the linear MR was attributed to the interaction between bulk and TSS [24–26].

The nonlinearity of the $R_{xy}$ curves in Fig. 3(b) indicates the presence of more than one type of carrier in the Sb films. When considering the band structure of Sb from previous studies [7], the Fermi level intersects both the conduction and valence bands in the Γ-M direction, as well as the surface states. We employed a 3-band model to obtain the carrier concentration and respective carrier mobility for each of the carriers—namely, those from the conduction band, valence band, and surface states [27]. Resulting carrier concentrations include one positive concentration and two negative concentrations, implying two electron carrier channels and one hole carrier channel (Details of the 3-band model fitting is provided in Supplemental Material with Fig. S2 showing a fitting on the 13.2 nm sample). Figures 3(c) and (d) show the variation of carrier concentration and carrier mobility for each Sb film. Absolute values of carrier concentration are plotted for a clear comparison. Examining the order of magnitude of the carrier concentration, one of the electron channels is significantly smaller than the other two. Additionally, carrier mobility of this electron channel appears to be greater than the mobilities of the others with higher carrier concentrations. Due to this discrepancy in both concentrations and mobilities, it is likely that the electron channel with smaller carrier concentration originates from the surface states, while the other two carrier channels originate from bulk bands, with the hole channel coming from the bulk valence band and the electron channel from the bulk conduction band. Furthermore, the electron and hole carrier concentrations from the bulk bands tend to decrease as the thickness of the Sb film decreases, which is consistent with the opening of a bandgap with a decrease in thickness.

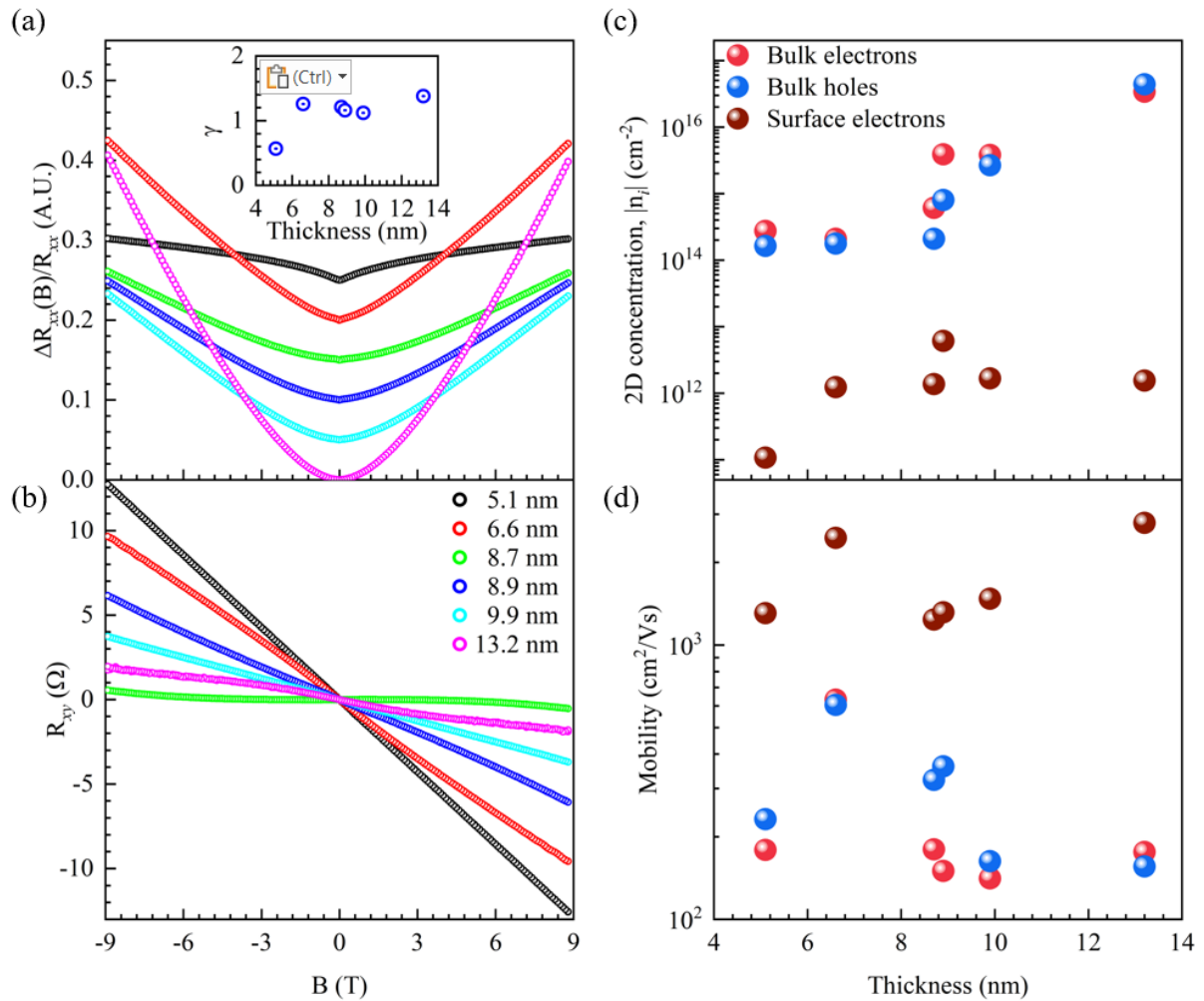

Figure 3. (a) Normalised MR at 4 K for Sb films of varying thicknesses. A vertical offset is applied for clarity to highlight the high-field MR behaviour. *Inset:* the variation of the exponent, γ with the thickness of the sample. (b) Transverse resistance curves showing non-linearity, consistent with the presence of multiple carrier types; the slope varies systematically with thickness. (c) and (d) Absolute carrier concentrations and mobilities, respectively, extracted using a three-band model. Both quantities exhibit an overall decreasing trend with reduced film thickness.

## C. Angle-Resolved Photoemission Spectroscopy

Figure 4 presents band dispersion from ARPES measurements for the two Sb thin films (15 nm and 6 nm) grown on $(Bi,Sb)_2Te_3$ buffer layers on sapphire substrates. Band dispersion along the Γ-M direction for the two films (Figs. 4(a) and 4(e)) shows the two surface states (the V-shaped features originating around -0.18 eV energy) and the bulk valence band below the surface states. Second-derivative plots of the band dispersion (Figs. 4(b) and 4(f)) focus on the surface states within a narrower energy range in the vicinity of the Fermi level. These plots clearly reveal the two surface states of Sb, indicated by the red dashed lines, consistent with prior ARPES studies of Sb [28,29]. The corresponding Fermi surfaces are displayed in Figs. 4(c) and 4(g) for the 15 nm and 6 nm films, respectively. The sixfold symmetry of the surface states at the Γ point is clearly resolved in the 15 nm film but appears less defined in the thinner 6 nm sample. Additionally, a feature at the M points, visible in the Fermi surface of the 15 nm film, is attributed to the bulk conduction band. This characteristic is notably absent in the 6 nm film, indicating a thickness-dependent change in electronic band structure. To further investigate the band dispersion at the M point, additional scans along the Γ-K direction were conducted along the blue lines in the Fermi surface plots, as shown in Figs. 4(d) and 4(h) for the 15 nm and 6 nm films, respectively. The thicker sample exhibits a high-intensity signal below the Fermi level, which is mostly absent in the 6 nm film. Previous ARPES studies on thick Sb films have indicated that the conduction band extends below the Fermi level at the M-point [30]. Hence, it is evident that the band observed for the 15 nm film at the M-point originates from the conduction band. The absence of this signal in the 6 nm film at the M-point strongly supports that with a decreasing Sb film thickness, there is a shifting of the conduction band relative to the Fermi level, suggesting that the conduction band has been lifted with the decrease in the film's thickness.

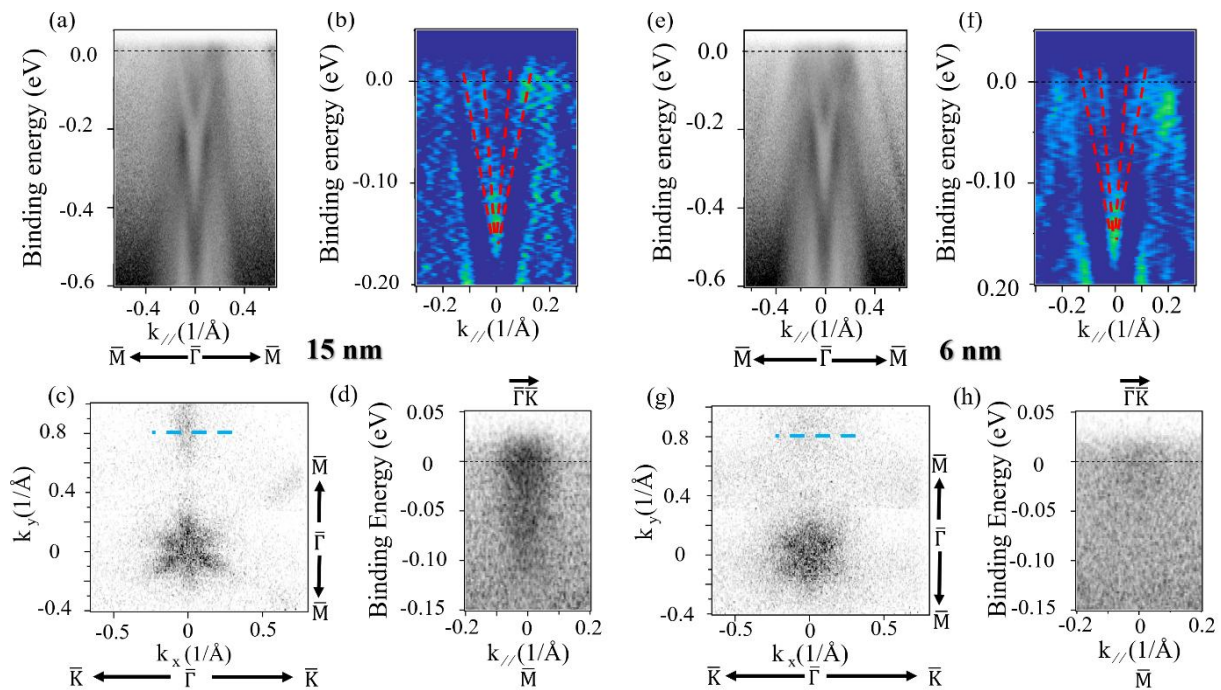

Figure 4. (a) Band dispersion in the Γ-M direction of the 15 nm film while (b) Second derivative plot of (a), the red dotted lines are overlaid on the surface states to enhance their visibility and aid in visualisation. (c) Fermi surface of the 15 nm film and (d) shows the ARPES scan performed in the Γ-K direction along the blue dotted line in (c). (e) Band dispersion in the Γ-M direction of the 6 nm film. (f) second derivative plot of (e) with the red dotted lines for enhanced visibility of the surface states. (g) Fermi surface of the 6 nm film and (h) shows the ARPES scan performed along the Γ-K direction along the blue dotted line in (g).

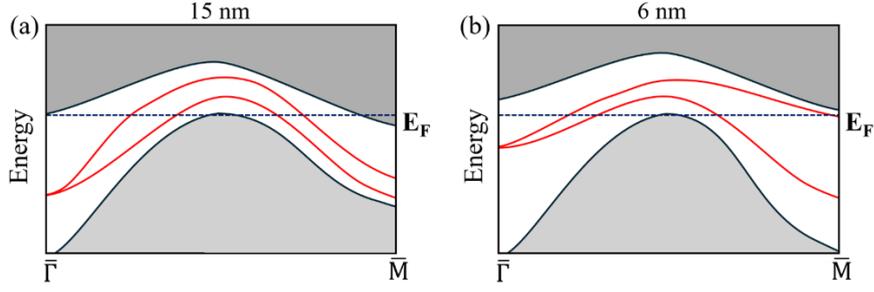

Figure 5. Shifting of bulk bands is illustrated with the thickness for (a) 6 nm and (b) 15 nm films, respectively; the red lines indicate the surface states, and the dashed lines indicate the Fermi level.

Our ARPES results agree with the theoretical work predicted by Zhang et al. [7], in which the conduction band at the M-point shifts upward with decreasing Sb film thickness, leading to the opening an indirect band gap in the material due to the quantum confinement effect [7]. To further clarify this interpretation, we have reconstructed the energy band dispersion diagrams as schematics in Figs. 5(a) and 5(b), along the Γ-M direction for the 15 nm and 6 nm films, respectively. These schematics illustrate the upward shift of the conduction band and the downward shift of the valence band with decrease in Sb thickness. These results are also consistent with our electrical transport measurements, which show that a reduction in both electron and hole concentrations with decreasing Sb thickness.

### D. Weak antilocalization

The normalized longitudinal MR curves for all the samples show a dip in resistance near zero magnetic field at 4 K, as shown in Fig. 6(a). The notable shift observed in MR can be elucidated through weak antilocalization (WAL) [31]. WAL is seen from materials which have strong spin-orbit coupling and from materials which have TSS [32]. In TSSs, WAL arises from the unique helical spin texture, of the massless Dirac fermions which have a $\pi$ Berry phase leading to quantum interference among electrons traversing pairs of time-reversed self-crossing trajectories, thereby suppressing backscattering [33]. This suppressed backscattering yields a positive adjustment to conductivity at low temperatures [24]. To accurately describe this negative magnetoconductivity, the Hikami–Larkin–Nagaoka formula is employed, which successfully accounts for first-order interference effects [34,35]:

$$\Delta\sigma(B) = -\alpha \frac{e^2}{2\pi^2\hbar} \left( \psi\left(\frac{1}{2} + \frac{B_\phi}{B}\right) - \ln\left(\frac{B_\phi}{B}\right) \right), \quad (1)$$

where e is the electronic charge, $\hbar$ is the Planck's constant, $B_\phi = \frac{\hbar}{4el_\phi^2}$ is the characteristic magnetic field with $l_\phi$ being the phase coherence length, and Ψ(x) is the digamma function. The prefactor $\alpha$ given in Eq. (1) gives an idea about the number of coherent conducting channels. An $\alpha$ value of -0.5 corresponds to one conducting channel and an $\alpha$ value of -1 corresponds to two conducting channels [32,36]. Upon close inspection of the range of WAL for the films at 4 K reveal that the range of the magnetic field in which the dip in resistance occurs depends on the sample considered. It is clearly seen that the magnetic field range in which WAL appears broadens with the decrease in the thickness of the film. It goes from about 0.1 T in the thickest sample to about 1 T in the thinnest sample implying that the magnetic field needed to destroy the destructive interference of time reversed electron pairs must be stronger with the reduction in thickness of the films. This type of a variation is also reported for Bi$_2$Se$_3$ thin films where the parabolic nature of the thicker films suggests a larger contribution from the bulk bands to transport relative to the TSS [37,38]. All the samples showed $\alpha$ values around -0.5 at 4 K when fitted with Eq. (1) with a suitable magnetic field range in which WAL occurs and is summarised in Fig. 6(b) along with $l_\phi$. (See Supplemental Material S3 for a representative WAL fitting for the 13.2 nm Sb film.) In 3D TIs with helical spin textures, this $\alpha$ value likely indicates the existence of a single coherent conducting channel, suggesting coupling between the top and bottom surface states via the bulk bands. This coupling gives rise to a single channel responsible for the quantum correction to conductivity near zero magnetic fields [37]. Moreover, WAL has also been reported in TSMs [39], such as Topological Dirac semimetal $Cd_3As_2$ [40–42] and topological Weyl semimetal TaAs with $\alpha$ being close to -0.5 [43].

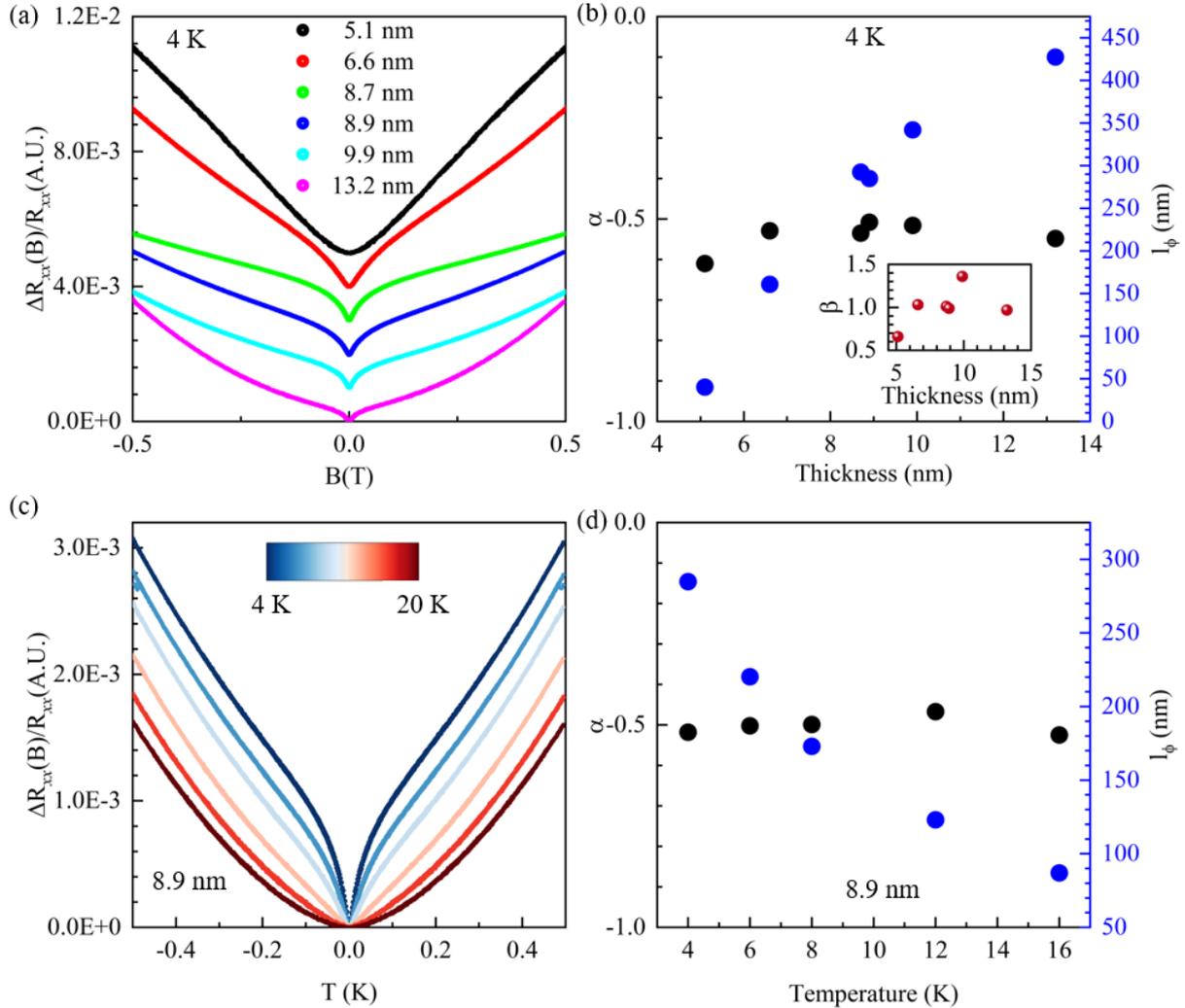

Figure 6. (a) Normalized longitudinal resistance of the films near zero magnetic field at 4 K showing that the range over which WAL exists increases with the decrease in thickness of the films. A vertical offset is applied for clarity to highlight the range of WAL. (b) $\alpha$ and $l_\phi$ values relating to the Sb thin films of different thickness at 4 K; $\alpha$ remains constant around -0.5 while $l_\phi$ drops with the decrease in thickness. (c) $R_{xx}$ vs B of the 8.9 nm film showing that WAL disappears with increasing temperature and becomes parabolic at 20 K. (d) The variation of ß, signifying the change in phase decoherence mechanism with thickness.

The parameter $l_\phi$ represents the distance over which an electron maintains its phase coherence. At 4 K, $l_\phi$ decreases with decreasing film thickness, following a sub-linear relationship with respect to the thickness (Figs. 6(a) and 6(b)). The nature of this thickness dependence provides insight into the contribution of the surface states to the WAL effect. In metals with trivial topology and strong spin-orbit coupling, $l_\phi$ is expected to scale linearly with thickness. In contrast, for ideal TIs with no bulk conduction, $l_\phi$ should remain independent of thickness. In intermediate cases, where non-trivial TSS coexist with bulk bands, $l_\phi$ typically exhibits a sublinear dependence of thickness [37]. The observed trend in our Sb films suggests the presence of non-trivial TSS alongside bulk conduction, consistent with previous reports on $Bi_2Se_3$ thin films [25].

Analysis of the WAL phenomenon across various temperatures in all the films, reveals that the phase coherence lengths decrease with increasing temperatures, while the alpha value remains close to -0.5 up to 16 K for the thicker samples and becomes zero at 20 K, at which point the MR curves become parabolic. Variation of longitudinal MR showing the change of WAL with the temperature for the 8.9 nm (as a representation) is shown in Fig. 6(c). The variation of the parameters alpha and $l_\phi$ with temperature is shown in Fig. 6(d). The decrease in phase coherence length with rising temperature is due to the rapid decoherence of electrons from increased electron-electron or electron-phonon interactions, which also results in the disappearance of the WAL behaviour. This trend of the decrease in $l_\phi$ with temperature is seen in all the samples.

The phase decoherence mechanism of each sample can be inferred by examining the variation of $l_\phi$ with temperature ($l_\phi \sim T^{-\beta}$). For 2D systems with electron-electron interactions $\beta=0.5$ and $\beta>1$ for electron-phonon interactions [44–47]. By plotting the variation of $l_\phi$ with respect to temperature with a power law relationship for each film separately, we were able to determine the exponent, $\beta$ (Inset in Fig. 6(b)). The exponent of the thinnest sample is -0.63 while all the other samples showed an exponent value larger and close to -1 indicating a change in the phase decoherence mechanism with the change in thickness of the films (The power-law fitting of $l_\phi$ versus temperature for the thinnest and thickest samples is shown in Fig. S4 of the supplementary section, from which the value of $\beta$ was extracted). Hence our results indicate that the thinnest sample shows 2D electron-electron dephasing and the dephasing mechanism in the rest of the films is 3D electron-phonon interactions. This shows that the with the decrease in the thickness of the films, the effect from surface states on the transport results become significant [48,49].

## IV. CONCLUSIONS

Through careful optimization of growth conditions, we successfully grew epitaxial Sb thin films along the (001) hexagonal direction on both GaSb(111)B and sapphire substrates. This enables us to systematically study changes in the band structure of Sb induced by the quantum confinement effect. Electrical transport measurements revealed that, as the Sb film thickness decreased, the temperature dependence of the sheet resistance transitioned from metallic to insulating behaviour. Transverse MR revealed a decrease in bulk band carrier concentrations with decreasing thickness, consistent with an opening of a band gap. This observation was further corroborated by ARPES, which showed an upward shift of the conduction band with decreasing thickness, indicating gap formation.

The appearance of WAL in the longitudinal MR for Sb films with the thickness ranging from 5.1 nm to 13.2 nm suggests a potential contribution from TSS. WAL analysis yielded the prefactor $\alpha$ = -0.5, indicating the presence of one coherent conducting channel. However, the presence of multiple carriers, as indicated by transverse MR measurements, suggests that bulk conduction significantly influences the transport results. This bulk conduction effect, likely due to the Fermi level being near or crossing the bulk bands depending on film thickness, complicates the understanding of the TSS and the topological nature of Sb thin films. To address this, future studies will need to employ electrostatic gating to tune the Fermi level into the bulk bandgap, thereby minimizing the influence of bulk carriers.

In summary, our electrical transport and ARPES measurements consistently demonstrate that reducing the thickness of Sb thin films leads to the upward and downward shifts of conduction and valence bands, respectively. These findings establish a solid foundation for exploring the non-trivial topology in Sb and experimentally realizing its predicted topological phase transition.

**ACKNOWLEDGMENTS**

This work was supported by the Science Alliance at the University of Tennessee, Knoxville, through the Support for Affiliated Research Teams (StART) program. Part of this study is based upon research conducted at The Pennsylvania State University Two-Dimensional Crystal Consortium – Materials Innovation Platform (2DCC-MIP) which is supported by NSF cooperative agreement DMR-2039351. P.A. acknowledges support from the U.S. Department of Energy, Office of Science, Basic Energy Sciences, Materials Sciences and Engineering Division, for the material growth efforts.

# Supplemental Material

## S1. X-ray reflectivity

X-ray reflectivity (XRR) measurements were performed to determine the thickness of the Sb thin films. Figure S1 presents the XRR data along with the corresponding fit for the thickest sample, yielding a thickness of 13.2 nm. This fitted thickness was used to estimate the growth rate of Sb on the GaSb(111)A substrate at the given growth temperature. The thicknesses of the other samples were determined in the same manner.

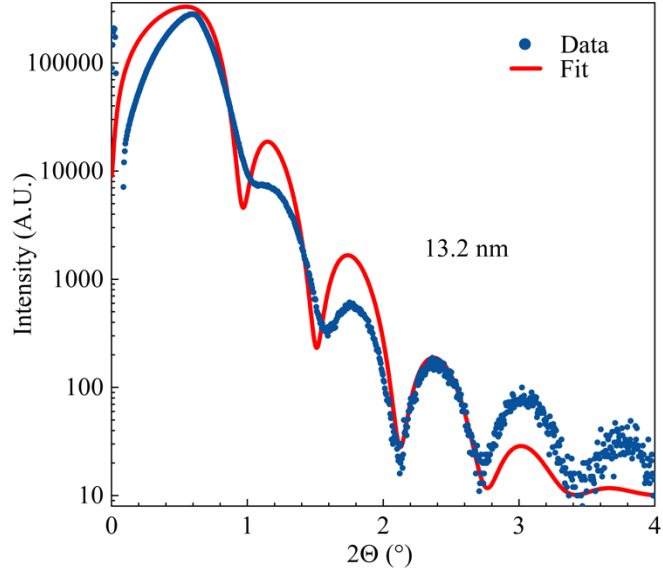

**Figure S1.** XRR fitting for the thickest Sb thin film yielding a thickness of 13.1 nm.

## S2. Triple-band Hall effect model

Due to the non-linearity observed in the transverse resistance, the Hall resistance data were fitted using a triple-band model to account for the presence of multiple carrier types [1]. The total Hall resistance is described as the superposition of three parallel conducting channels, each associated with a carrier type characterised by its own mobility and carrier concentration. The equation used for the fitting is provided below:

$$\sigma_{xy} = \sum_{i=1}^{3} \frac{n_i e \mu_i^2 \mu_0 B}{1+(\mu_i \mu_0 H)^2} \quad \text{(S1)}$$

Here, $n_i$ and $\mu_i$ represent the two-dimensional carrier density and the mobility of the $i$-th channel, respectively. $\mu_0$ is the permeability of free space, $B$ is the magnetic field, and $e$ is the elementary charge. The two-dimensional transverse sheet conductivity $\sigma_{xy}$, is defined as:

$$\sigma_{xy} = \frac{R_{\square\_xy}}{R_{\square\_xx}^2 + R_{\square\_xy}^2} \quad \text{(S2)}$$

Here $R_{\square\_xx} = R_{xx}\frac{w}{l}$ and $R_{\square\_xy} = R_{xy}$ (w, and l represent the width and length of the Hall bar channel respectively.) $R_{\square\_xx}$ and $R_{\square\_xy}$ are the longitudinal and transverse sheet resistances respectively derived from the longitudinal resistance, $R_{xx}$ and Hall resistance, $R_{xy}$. Figure S2 shows the fitting for the 13.2 nm film using equation (S1). The same fitting procedure was applied to extract the carrier concentrations and mobilities for the other samples. It should be emphasised that the fittings are performed only at low magnetic fields, since Equations S1 and S2 are applicable only in a small-field regime.

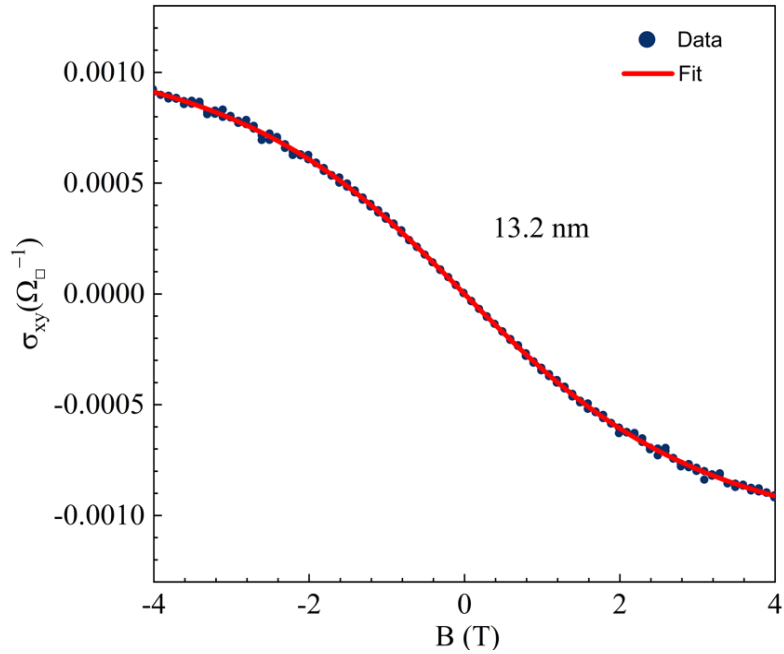

**Figure S2.** The triple-band Hall fitting for 13.2 nm sample.

**S3. Weak antilocalization**

Weak antilocalization (WAL) effects were analysed using the Hikami–Larkin–Nagaoka (HLN) model [2]. To obtain the low-temperature conductivity correction in equation (1) of the Results and Discussion section, the longitudinal sheet conductivity, $\sigma_{xx}$, was calculated as shown in equation S(3) below.

$$\sigma_{xx} = \frac{R_{\square\_xx}}{R_{\square\_xx}^2 + R_{\square\_xy}^2} \tag{S3}$$

The HLN fitting for the 13.2 nm sample at 4 K is shown in Figure S3. The same method was used to extract the HLN parameters for all other samples.

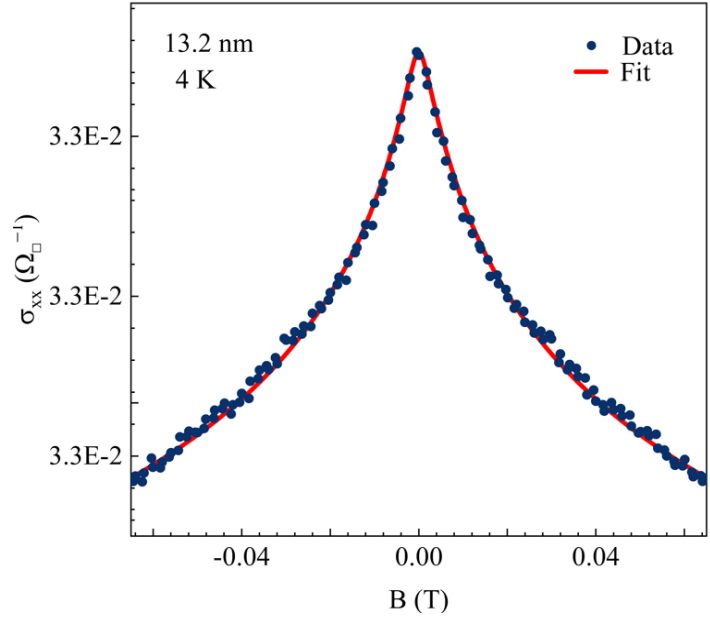

**Figure S3.** WAL fitting for the 13.2 nm sample at 4 K.

**S4. Temperature dependence of phase coherence length**

The temperature dependence of $l_\varnothing$ was fitted using a $l_\varnothing = T^{-\beta}$ relationship [3] [4]. Figure S4 compares this behaviour for the 13.2 nm and 5.1 nm samples. The ß values for all other samples were also determined using the same approach.

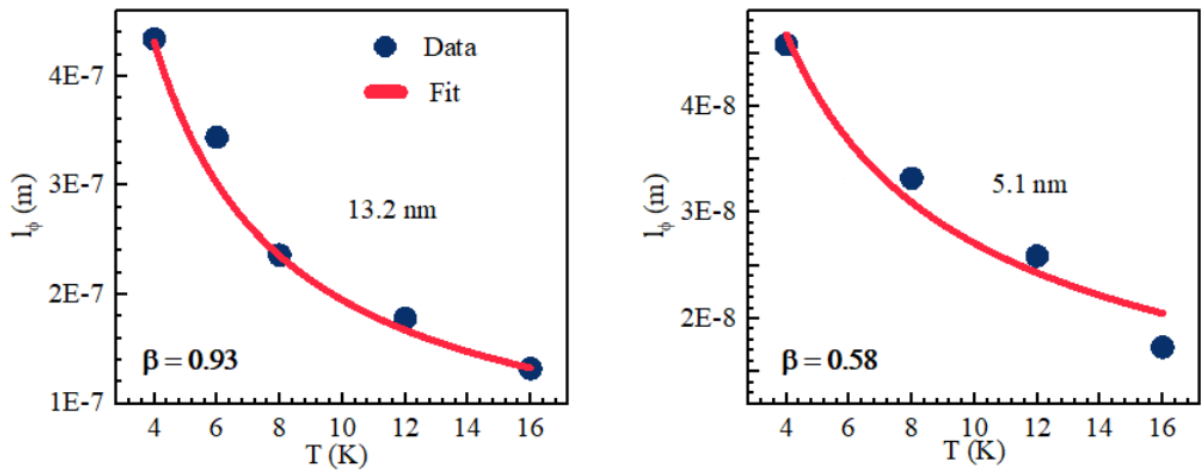

**Figure S4.** Temperature dependence of phase coherence length.